\documentclass[]{article} 
\usepackage{times}  
\usepackage{helvet}  
\usepackage{courier}  
\usepackage[hyphens]{url}  
\usepackage{graphicx} 
\usepackage{caption} 
\usepackage{amsmath}
\usepackage{authblk}

\usepackage[numbers]{natbib}

\usepackage{amssymb}
\usepackage{algorithm}
\usepackage{algorithmic}

\usepackage{newfloat}
\usepackage{listings}

\floatstyle{ruled}
\newfloat{listing}{tb}{lst}{}
\floatname{listing}{Listing}
\title{A Semi-Parametric Model for Decision Making in High-Dimensional Sensory Discrimination Tasks}

\author[1,2]{Stephen Keeley}
\author[2]{Benjamin Letham} 
\author[2]{Chase Tymms} 
\author[2]{Craig Sanders} 
\author[2]{Michael Shvartsman}

\affil[1]{Department of Natural Sciences, Fordham University, USA} 
\affil[2]{Meta}

\begin{document}

\maketitle

\begin{abstract}
Psychometric functions typically characterize binary sensory decisions along a single stimulus dimension. However, real-life sensory tasks vary along a greater variety of dimensions (e.g.\ color, contrast and luminance for visual stimuli). Approaches to characterizing high-dimensional sensory spaces either require strong parametric assumptions about these additional contextual dimensions, or fail to leverage known properties of classical psychometric curves. We overcome both limitations by introducing a semi-parametric model of sensory discrimination that applies traditional psychophysical models along a stimulus intensity dimension, but puts Gaussian process (GP) priors on the parameters of these models with respect to the remaining dimensions. By combining the flexiblity of the GP with the deep literature on parametric psychophysics, our semi-parametric models achieve good performance with much less data than baselines on both synthetic and real-world high-dimensional psychophysics datasets. We additionally show strong performance in a Bayesian active learning setting, and present a novel active learning paradigm for the semi-parametric model.
\end{abstract}

\section{Introduction}

Understanding the mappings from physical stimuli to mental percepts is an important goal of perceptual neuroscience and psychophysics. A popular experimental technique is to measure behavioral responses to varying levels of a single stimulus feature, such as contrast of an image or volume of a sound, and generate a behavioral response curve. The average response probabilities are then often fit to a functional form that is sigmoidal 
in shape, for example using the probit, logit, or Weibull functions \cite{strasburger2001converting}. These parametric models tend to have specific interpretable parameters that are of use to practitioners, such as thresholds and slopes  \cite{brand2002efficient}. 
Improving the accuracy and sample efficiency of parameter estimation for these psychometric functions remains an area of active research \cite{schutt2016painfree, shen2012maximum}. However, much of the existing work focuses on a single stimulus feature, ignoring the fact that stimuli continuously vary in important dimensions other than intensity, such as color or pitch. To understand sensory sensitivity in these settings, univariate parameterized models require densely sampling the entire stimulus feature domain; i.e. creating a single psychometric function for each value of any non-intensity stimulus feature. 

Recent work overcomes these limitations and extends the classical densely-sampled univariate psychophysical curve in terms of both modeling and experimental design. On the modeling front, recent work captures correlations in psychometric functions across stimulus dimensions, either using prespecified multidimensional parametric models \cite{watson2017quest+}, or using nonparametric Gaussian processes (GPs) \cite{gardner2015bayesian, gardner2015psychophysical, owen2021adaptive}. For experimental design, dense sampling of the entire multidimensional input space has been replaced with efficient active learning schemes \cite{houlsby2011bayesian, settles2009active, gardner2015bayesian,gardner2015psychophysical, watson2017quest+}. Together, flexible psychometric models and multidimensional adaptive sampling methods have led to improvements in estimating psychometric tuning in multivariate stimulus settings \cite{gardner2015bayesian,owen2021adaptive, letham2022look}.

However, existing high-dimensional psychometric models suffer from some limitations. The fully parametric approach, while interpretable, is constrained in practice, requiring an a priori model for all dimensions, with parameters learned using an inefficient grid-search \cite{watson2017quest+}. On the other hand, the more flexible, nonparametric models that leverage the power of GPs are not regularized to have sensible tuning properties, such as positive monotonicity in stimulus intensity. Moreover, they do not afford an experimenter with interpretable parameters like those given in the classical univariate context \cite{gardner2015psychophysical,owen2021adaptive}.

To address these limitations, we propose a semi-parametric model of the psychometric field well-suited for high-dimensional stimuli. Over the intensity dimension, our psychometric model is governed by a characteristic function that has a sigmoidal shape with identifiable slope and offset parameters. Each of these parameters, however, is governed by a GP across context (non-intensity) stimulus dimensions, admitting a flexible characterization of the psychometric field in a high-dimensional continuous stimulus space. While the posterior of our model is intractable, we develop two approximations, each with different benefits. The first, the \textit{full semi-parametric model}, uses a semi-parametric variational posterior that factorizes over the slope and offset parameters of the sigmoid to learn the model. For the second, the \textit{MVN approximate model}, we derive a new approximation for the elementwise product of the multivariate normal (MVN) distributions of the slope and offset GPs in our model. This approximation implicitly defines a new, single GP kernel specific to the psychophysics setting, and lets us perform inference with fewer variational parameters, maintaining a form that is adaptable to standard GP inference methods and active learning machinery. We evaluate our model on simulated and real data in up to 8 dimensions, and find that this semi-parametric approach not only provides interpretable results in high-dimensional stimulus settings, but also offers a faster and more accurate estimation procedure for the psychometric field

Given that Bayesian active learning also plays an important role in high dimensional psychophysical tuning estimation \cite{gardner2015bayesian,owen2021adaptive, letham2022look}, we conclude our model evaluation showing semi-parametric performance under a variety of existing active learning objectives. We find that our full semi-parametric model shows strong performance under a variety of these objectives, and our MVN approximate model allows for analytic acquisition functions, including recently-developed look-ahead approaches \cite{letham2022look}. Finally, we introduce a novel threshold-based acquisition function for use with our full semi-parametric model that shows strong active learning performance in a high-dimensional experiment. 

\begin{figure*}
\centering
\includegraphics[width = 1\textwidth]{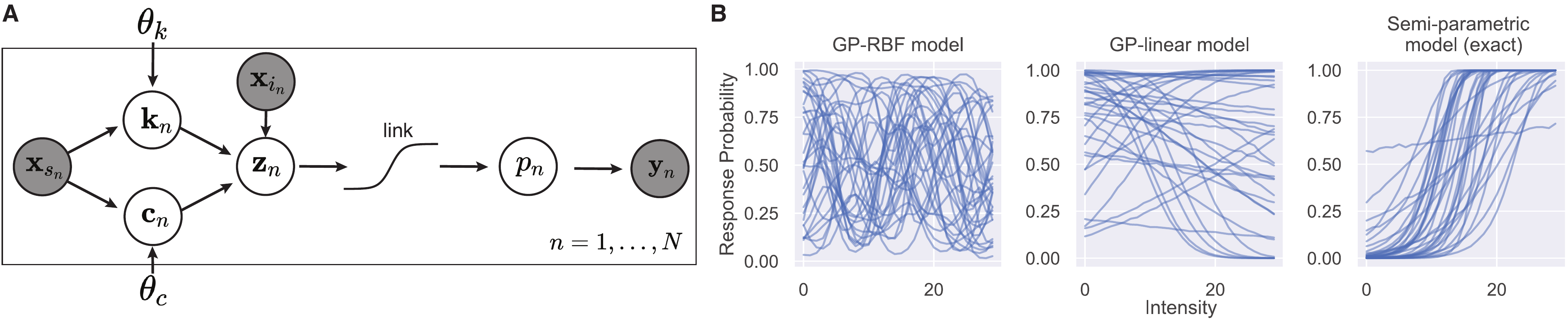}
\caption{\textbf{A.} Graphical model of our semi-parametric approach. \textbf{B.} Samples from the prior along the intensity dimension for three psychophysical models. The GP-RBF model's prior contains unrealistic psychometric functions that are not monotonic. 
The previously published GP-linear model 
overly restricts the shape of the sigmoid due to the linear kernel in intensity, while admitting functions with negative slopes. Our new model's prior, here with a Gumbel link, contains a more diverse set of realistic psychometric functions than either baseline variant.}
\label{fig:graphical}
\end{figure*}
\subsection{The Semi-Parametric Psychophysical Model}

We consider data of the form $D = \{\mathbf{x}_n,y_n\}_{n=1}^N$ where $y_n \in \{0, 1\}$ are participant responses and $\mathbf{x}_n = (x_{i_n}, \mathbf{x}_{s_n})$ describe stimulus configurations. We separate stimulus context ($s$) dimensions from the intensity ($i$) dimension. Throughout the manuscript we only consider 1-dimensional intensity, so that $x_{i_n}$ is a scalar. We assume that intensity predicts responses according to a standard psychophysical parametric model of the form $p = \sigma(k (x_i + c))$ for some slope $k$, offset $c$, and sigmoid link $\sigma: \mathbb{R} \rightarrow [0, 1]$ and probability of detection $p$. Our framework can be flexibly adapted to incorporate any sigmoid from the literature, including the probit (as standard in GP classification), logistic (common elsewhere in machine learning), Weibull or Gumbell CDF (common in psychophysics, \cite{strasburger2001converting, may2013four}), or other common probability mappings. We can additionally modify these sigmoids, for example shifting or scaling them to manage psychophysics experiments where participants are asked to discriminate between stimuli, and the response probability is lower-bounded above 0. Common experimental paradigms can produce a lower bound response probability at 0.5 (for two-alternative decisions) or 0.25 (for four-alternative `odd one out' trials).

We place independent GP priors on the slope and offset governed by the context dimensions, which implies that each of $\mathbf{f}_k = [k(\mathbf{x}_{s_1}), k(\mathbf{x}_{s_2}), \ldots,  k(\mathbf{x}_{s_N})]$ and $\mathbf{f}_c = [c(\mathbf{x}_{s_1}), c(\mathbf{x}_{s_2}), \ldots, c(\mathbf{x}_{s_N})]$ is Gaussian distributed: 
\begin{equation}\label{eq:gaussians}
\mathbf{f}_k \sim \mathcal{N}\left(m, \Sigma_{k}\right), \mathbf{f}_c \sim \mathcal{N}\left(0, \Sigma_{c}\right),    
\end{equation}
where $\Sigma_{k},\Sigma_{c}$ are $N \times N$ covariance matrices whose $(n,n')$th entry is given by a kernel function $\kappa(\mathbf{x}_{s_n},\mathbf{x}_{s_{n'}})$ and $N$ is the total number of stimuli sampled.  We use the standard radial basis function (RBF) kernel with independent hyperparameters for the slope and offset GP kernels. Each kernel is governed by its own hyperparameters $\theta_c$ and $\theta_k$ for the offset and slope GPs, respectively. Finally, $m$ is a positive constant to center the prior distribution of slope values at some positive number (we use $m=2$ for all experiments; see appendix for additional detail and evaluation).

In this formulation, we can write the joint distribution of latent stimulus values as: 
\begin{equation}
\mathbf{z} = \mathbf{f}_k \circ \left(\mathbf{f}_c + \mathbf{x}_{i}\right),
\end{equation}
where $\circ$ denotes the Hadamard (elementwise) product and $\mathbf{x}_i = (x_{i_1}, x_{i_2}, \dots x_{i_N}$). The likelihood of a set of observations is given by a set of independent Bernoulli distributions with probabilities equal to $\mathbf{z}$ taken through the sigmoid link. To make sense of this mapping, consider a joint draw of $N$ slopes $\mathbf{f}_k \in \mathbb{R}^N$ and intercepts $\mathbf{f}_c \in \mathbb{R}^N$. Suppressing the dependence of $k$ and $c$ on $\mathbf{x}_s$, the resultant transformed collection of variables has each $z_n =k_n c_n + k_nx_{i_n}$, producing $\mathbf{z}  = [k_1c_1 + k_1x_{i_1}, k_2c_2 + k_2x_{i_2}, \ldots, k_N c_N + k_Nx_{i_N}]$. Note that this is just a multivariate extension of the standard $k(x + c)$ input into the link, written jointly to accommodate the GP prior on $k$ and $c$. A graphical depiction of the semi-parametric model is given in Fig.~\ref{fig:graphical}\textbf{A}.

Fig.~\ref{fig:graphical}\textbf{B} shows prior samples of the probability of detection, $p$, plotted along the intensity dimension, for three different models. The first is an unconstrained GP model (GP-RBF) which treats all dimensions equally \cite{owen2021adaptive}, and the second has an additive GP kernel that is linear in the intensity dimension and RBF in context dimensions (GP-linear) \cite{gardner2015psychophysical}. The third (Fig.~\ref{fig:graphical}\textbf{B}, Right) is our proposed semi-parametric model. These probability samples for each model are drawn from the prior and illustrate the hypothesis space of the considered psychometric functions. The GP-RBF model allows for tuning that is not sigmoidal (or even monotonic), and the GP-linear model restricts the shape of the sigmoid while still permitting negative slopes. In contrast, our semi-parametric model shows classical psychometric curves along the intensity dimension and is free to vary independently in its offset $c$ and slope $k$. 

Our semi-parametric model is easily adaptable to use any link function with interpretable slope and intercept values, as well as scaled and/or shifted variants thereof. In the present contribution we consider the logistic and Probit mapping, as well as the so-called Weibull function which is the cdf of a left Gumbel distribution in log-space \cite{strasburger2001converting, may2013four}. We also consider a `floor link' whose minimum is set to the known lower bound of the response probability in the given task. For our results on simulated functions, we choose the link that performed best with quasi-random sampling for each model\footnote{This is the Gumbel-link with floor for the semi-parametric model for all except for GlobalMI acquisition, which requires a probit link with a floor of 0. For all other models this is the Probit link with a floor of 0.}. We show results with all links on human psychophysical data and include results from all links and floors in the appendix.

\section{Inference for the Semi-Parametric Psychophysical model}

The marginal likelihood for our model is: 

\begin{equation}
\begin{split}
P(&\mathbf{y}| \mathbf{X}_s,\mathbf{x}_i,\theta_k, \theta_c) = \\ &\int{p(\mathbf{y}|\mathbf{x}_i,\mathbf{f}_k,\mathbf{f}_c) p(\mathbf{f}_k|\mathbf{X}_s, \theta_k)p(\mathbf{f}_c|\mathbf{X}_s, \theta_c)d\mathbf{f}_k d\mathbf{f}_c}, 
\end{split}
\end{equation}
where $\mathbf{X}_s$ is a concatenation of $(\mathbf{x}_{s, 1}, \mathbf{x}_{s, 2} \dots \mathbf{x}_{s, N})$ and $\mathbf{y}$ is a concatenation of the $(y_1, y_2 \dots y_n)$ observations. This likelihood is intractable, but we provide two distinct strategies for approximating it: one by factorized variational inference (VI), and the other by approximating the model itself, which lets us apply standard VI methods for GPs. 

\subsection{Factorized Variational Inference for the Semi-Parametric Model}

We define MVN variational distributions $q_k$ and $q_c$ for the slope and offset to perform VI. We can write an evidence lower bound (ELBO) as follows:
\begin{align*}
\log p(\mathbf{y} \mid & \mathbf{X_s}, \mathbf{x}_i)  \geq \\ 
&\mathcal{L}:= \mathbb{E}_{q_c(\mathbf{f}_c), q_k(\mathbf{f}_k)}[\log p(\mathbf{y} \mid \mathbf{f}_k, \mathbf{f}_c,  \mathbf{x}_i)] \\
&\quad\quad-\textrm{KL}[q_k(\mathbf{f}_k) \| p(\mathbf{f}_k \mid \mathbf{X}_s)] \\
&\quad\quad- \textrm{KL}[q_c(\mathbf{f}_c) \| p(\mathbf{f}_c \mid \mathbf{X}_s)], 
\end{align*}
where the two KL terms between MVNs are available in closed form. For the remaining term, since the slope and offset GPs are combined into a scalar latent term per observation, we can compute the expectation by one-dimensional Gauss-Hermite quadrature. Gradients of all terms are available by automatic differentiation, which lets us optimize this objective by standard methods \citep{Hensman2015b,balandat2020botorch}. 
\subsection{Approximation to the Semi-Parametric Model}

While the inference approach above is tractable, its key disadvantage is that the prior on the latent $\mathbf{z}$ is no longer a Gaussian Process prior. This limits our ability to use standard black box GP variational inference for this model, as well as other tooling that relies on an MVN prior. 
Furthermore, since we have MVN posteriors on both slope and offset, the variational approximation has twice as many parameters, which potentially slows down inference (of interest for human-in-the-loop applications). To address these disadvantages, we derive an approximation to our model that avoids the need to use a variational approximation for both the slope and offset GP function values. Specifically, we directly approximate the latent function $\mathbf{z}$ with an MVN using moment-matching. 

We are interested in a MVN approximation to the latent function:
\begin{equation*}
    Z = F_k \circ \left(F_c + \mathbf{x}_{i}\right).
\end{equation*}
Here, $F_k$, $F_c$ are random variables distributed according to the Gaussians in (\ref{eq:gaussians}), and $Z$ is now also a random variable (in contrast to $\mathbf{f}_k$,$\mathbf{f}_c$, $\mathbf{z}$, indicating realizations of the random variables).  For convenience of derivation, let $\tilde{F}_c = F_c + \mathbf{x}_{i}$. The GP prior on $F_c$ implies that $\tilde{F}_c \sim \mathcal{N}\left(\mathbf{x}_{i}, \Sigma_{c}\right)$. We can compute the mean of $Z$ in terms of the slope and offset distributions:
\begin{equation*}
    \mathbb{E}[Z] = \mathbb{E}[F_k \circ \tilde{F}_c] = m \mathbf{x}_{i}.
\end{equation*}
We can similarly compute the covariance of the latent $Z$ in terms of the slope and offset. For convenience, denote $Y_k = F_k - m$ and $Y_c = \tilde{F_c} - \mathbf{x}_{i}$. We have that
\begin{align}\nonumber
    Z &= (Y_k + m) \circ (Y_c +  \mathbf{x}_{i})\\\label{eq:Z_eqn}
    &= m \mathbf{x}_{i} + m Y_c + \mathbf{x}_{i} \circ Y_k + Y_k \circ Y_c,
\end{align}
with $Y_c \sim \mathcal{N}(0,\Sigma_c)$ and $Y_k \sim \mathcal{N}(0,\Sigma_k)$. The covariance of the cross-term can be computed as
\begin{align}\label{crosscov_ln1}
    \textrm{Cov}[Y_k \circ Y_c]_{i, j} &= \mathbb{E}[(Y_k \circ Y_c)_i (Y_k \circ Y_c)_j]\\\nonumber
    &= \mathbb{E}[(Y_{k,i} Y_{c,i}) (Y_{k,j} Y_{c,j})]\\\label{crosscov_ln2}
    &= \mathbb{E}[Y_{k,i}Y_{k,j}] \mathbb{E}[Y_{c,i}Y_{c,j}]\\\nonumber
    &= (\Sigma_k)_{i, j} (\Sigma_c)_{i, j},
\end{align}
where (\ref{crosscov_ln1}) uses that $Y_k$ and $Y_c$ have 0 mean, and (\ref{crosscov_ln2}) uses their independence. Thus, $\textrm{Cov}[Y_k \circ Y_c] = \Sigma_k \circ \Sigma_c.$

Applying this result to (\ref{eq:Z_eqn}) we can compute the covariance for the latent function:
\begin{align*}
    \textrm{Cov}[Z] &= m^2 \Sigma_c + \mathbf{x}_{i} \mathbf{x}_{i}^{T} \circ \Sigma_k + \Sigma_c \circ \Sigma_k\\
    &= m^2 \Sigma_c + (\Sigma_c + \mathbf{x}_{i} \mathbf{x}_{i}^{T}) \circ \Sigma_k.
\end{align*}

The primary benefit of this approximation is that rather than learn the full semi-parametric latent function $Z$, which includes parameterizing slopes, $\mathbf{f}_k$, and offsets, $\mathbf{f}_c$, we simply infer values for the moment-matched approximate latent
\begin{equation}\label{eq:hadamard_approx}
\tilde{Z} \sim \mathcal{N}(m \mathbf{x}_{i}, m^2 \Sigma_c + (\Sigma_c + \mathbf{x}_{i} \mathbf{x}_{i}^{T}) \circ \Sigma_k).
\end{equation}

In this formulation, the latent function is effectively a GP with a novel kernel function specific to the psychophysics problem, and we can thus apply standard methods for model fitting, as well as for using the model in active learning as we will see below.

\subsubsection{Approximate Normality of the Latent Function}

The accuracy of the moment-matched MVN in (\ref{eq:hadamard_approx}) will depend on how close the true posterior for the latent function $Z$ is to being normally distributed. We show here the conditions under which $Z$ is Gaussian \cite{mathof_normallimit}.

Let $\Sigma = m^2 \Sigma_c + \mathbf{x}_{i} \mathbf{x}_{i}^{T} \circ \Sigma_k$. Then,
\begin{equation*}
    \Sigma^{-1/2}Z = W + \Sigma^{-1/2}(Y_k \circ Y_c),
\end{equation*}
where $W \sim \mathcal{N}(0,1)$.
Note that 
\begin{align*}
    \mathbb{E}\|\Sigma^{-1/2}(Y_k \circ Y_c)\|^{2} &\leq\left\|\Sigma^{-1 / 2}\right\|^{2} \mathbb{E}\left\|Y_k \circ Y_c\right\|^{2}\\
    &=\left\|\Sigma^{-1}\right\| \operatorname{tr}\left(\Sigma_{k} \circ \Sigma_{c}\right).
\end{align*}
Thus, as $\left\|\Sigma^{-1}\right\| \operatorname{tr}\left(\Sigma_{k} \circ \Sigma_{c}\right) \rightarrow 0$, $Z$ is Gaussian with mean $0$ and covariance $\Sigma$. This means that when the variances of the slope and offset values, $k$ and $c$, are low, the Hadamard product of the GPs is approximately Gaussian, and there will be little loss to using the Hadamard approximation model. Empirically, we later show that the Hadamard model incurs a modest performance loss relative to the full semi-parametric model, but (as shown in the appendix) the Hadamard model is much less able to benefit from psychophysics-specific modifications to the link function.

\begin{figure*}[t]
\centering
\includegraphics[width = 1\textwidth]{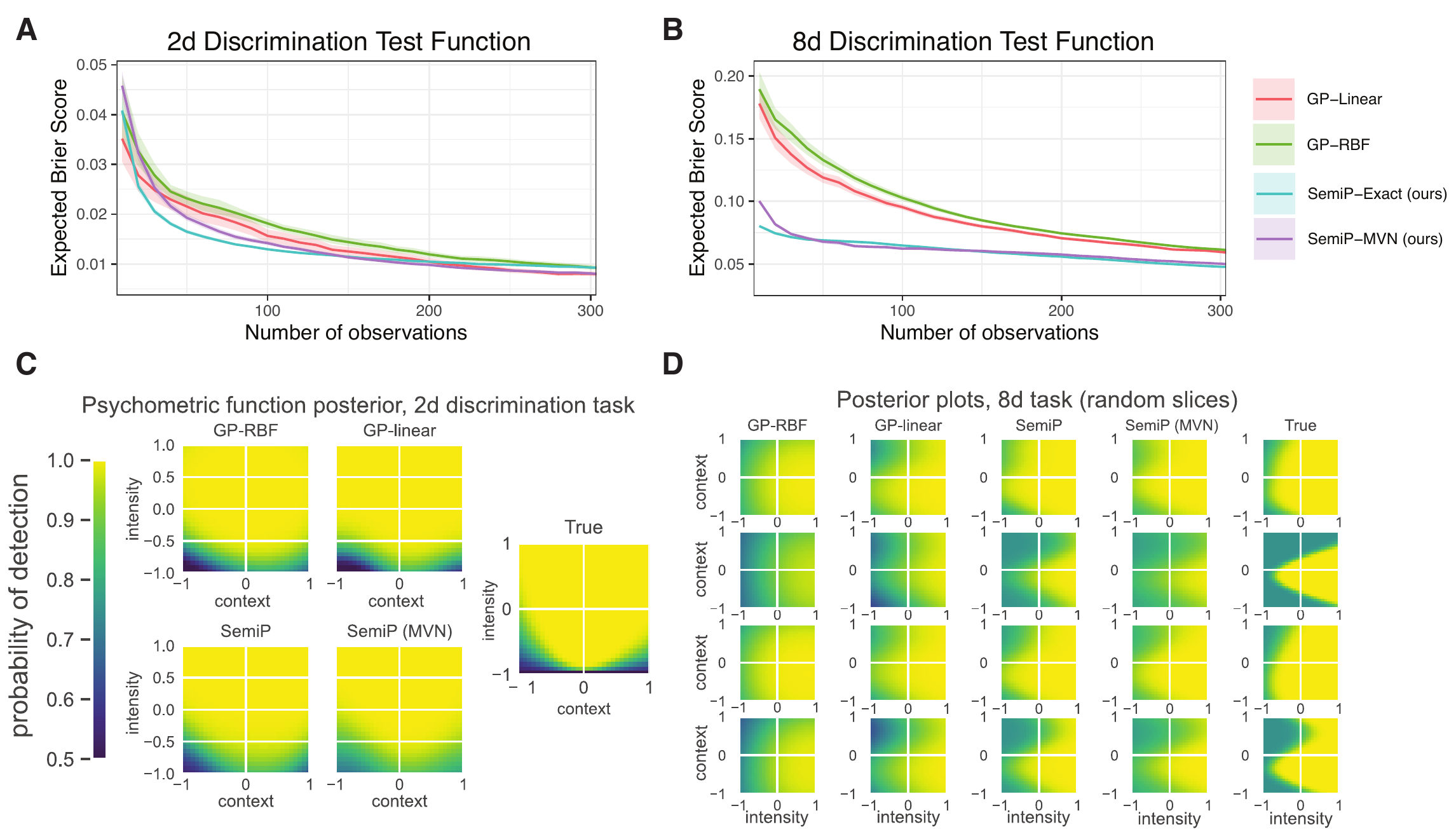}
\caption{\textbf{A.} Performance on 2-d example as a function of number of quasi-random samples drawn, for all models. Curves are averages from 100 replications, with standard errors shaded. \textbf{B}. Same for 8-d test function. 
\textbf{C.} 2-d function estimation after 500 samples for all models.  All models recover the test function with sufficient data.  
\textbf{D}. Inferred response probabilities in 8-d test function after 500 observations. Plots in each row have a randomly chosen dimension plotted along the y axis, and the intensity dimension along the x axis.}
\label{fig:2d_eval}
\end{figure*}






\section{Active learning with Semi-Parametric GPs}

Our semi-parametric model grants additional benefits for Bayesian active learning. Active learning methods define an acquisition function that prescribes the value of sampling a particular candidate point given the previously observed data and the current estimate of the model posterior. By optimizing this acquisition function, the next input ($\mathbf{x}$) is chosen for sampling. A key goal in psychophysics is the estimation of psychometric detection thresholds, more formally known as a \emph{level set estimation} (LSE) problem, i.e.\ finding the regions where the psychometric function is above or below some pre-determined threshold value $r$. In the general case of active learning for LSE, the location of the threshold is defined implicitly, and sampling strategies operate on an estimate of an arbitrary latent function. 

With our semi-parametric model, the threshold is uniquely determined as a function of the context dimensions, and can be computed directly from the model posterior for $k$ and $c$ as: 
\begin{equation*}
    x^r_i = \frac{\sigma^{-1}(r)}{k(\mathbf{x_s})}-c(\mathbf{x_s}). 
\end{equation*}
We can use samples from this posterior to compute several quantities of interest, for example the threshold posterior variance. In addition to being of interest to practitioners, reducing the posterior variance of the threshold is a natural, simple objective for active learning. We term it ThresholdBALV as it applies Bayesian Active Learning by Variance \cite{settles2009active} to the threshold posterior. 



 
\section{Results}

We showcase the benefits of the semi-parametric model in a few ways. First, we demonstrate performance on two synthetic psychometric test functions, where we show that our model can achieve good performance with less data than previously proposed baselines. Second, we evaluate performance on multiple real-world datasets, and show that our models outperform baselines in terms of predictive performance on unseen data. Finally, we demonstrate the compatibility of our models with active learning methods, again showing good performance with far less data than baselines, as well as competitive behavior of our novel ThresholdBALV acquisition function. We consider as baselines previous models used for flexible modeling and active learning for psychophysics, namely an otherwise-unconstrained GP-RBF model \citep{owen2021adaptive}, and a GP-linear model with a linear kernel in the intensity dimension and an RBF kernel in the remaining dimensions \citep{Schlittenlacher2018a,Schlittenlacher2020a, Song2017b,gardner2015psychophysical}. 


\subsection{Two dimensional task}

Before evaluating the semi-parametric model in a high dimensional setting, we first demonstrate performance in a simple 2-d psychometric test function, previously proposed in \citep{owen2021adaptive}, and detailed in the appendix. This function has a monotonic (probit) probability of detection along an intensity dimension, and smoothly varies as a linear combination of sines and cosines in a second dimension. 
We used a quasi-random Sobol sequence \cite{sobol1967distribution} to select stimulus locations $\mathbf{x}$ for our 2d test function. 

Fig.~\ref{fig:2d_eval}\textbf{A} shows the prediction performance of the semi-parametric model on this function and  Fig.~\ref{fig:2d_eval}\textbf{C} shows better estimation of the psychometric curve after 500 samples. For our evaluation metric we use the Brier score \citep{brier1950verification}, computed in expectation over the model's posterior. We use the Brier score because it assesses the calibration of the approximate posterior, and we use the expectation to account for the quality of posterior uncertainty estimation (we consider other metrics in the appendix). Our proposed model, with or without the MVN approximation, achieves low Brier scores much faster than baselines, though all models eventually achieve very good performance in this relatively simple test function. We will see next that the benefits are magnified in higher dimensions. 




\begin{figure*}
\centering
\includegraphics[width = 1\textwidth]{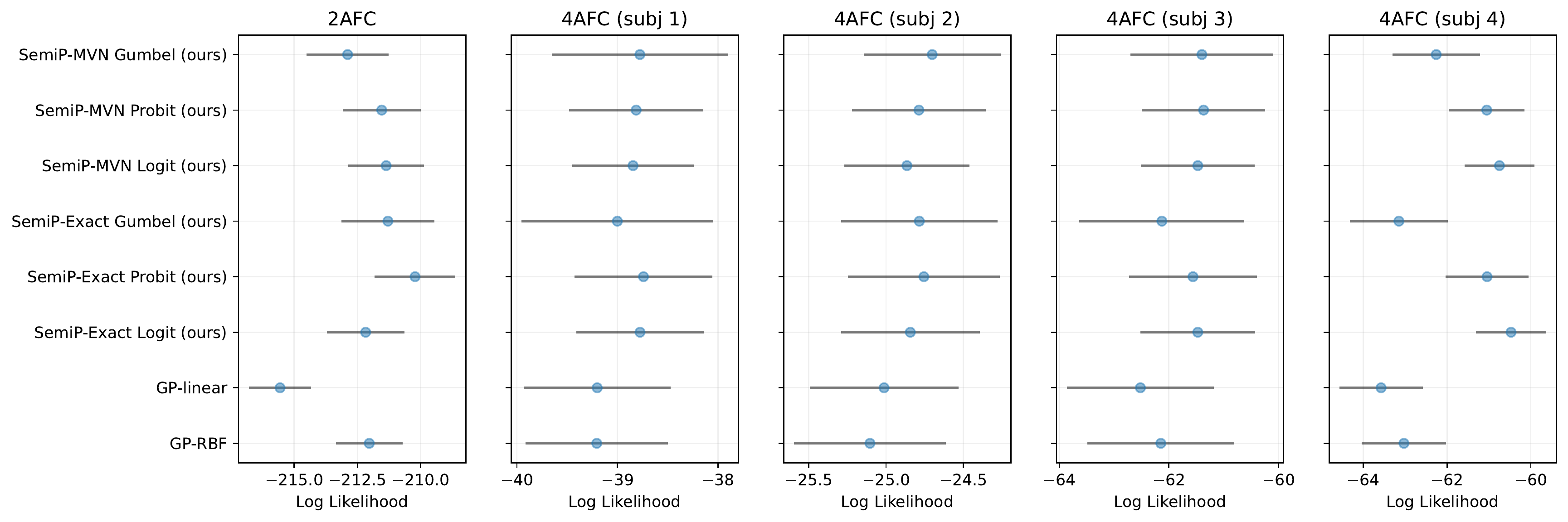}
\caption{Mean cross-validated log-likelihood for 15-fold cross validation on an example subject from a 2AFC psychometric task, and four example subjects in a 4AFC task from \cite{watson2017quest+}. Gray bars indicate standard error across folds.
}
\label{fig:xval}
\end{figure*}

\subsection{Eight dimensional task}

To simulate more a realistic, multidimensional sensory context we test the semi-parametric model using an 8-d psychometric test function previously reported by \citep{letham2022look} and described in the supplement. This function retains monotonicity in detection probability along the intensity dimension in the same way as the 2d function, but detection probability is a smooth function of stimulus feature inputs in the other seven dimensions. 

For this high dimensional test function, the semi-parametric model better estimates detection probability $p$ in fewer samples than competing models, and the approximate MVN model achieves essentially identical performance as the full semi-parametric model. Fig \ref{fig:2d_eval} \textbf{B} shows the Brier score for each model for the first 300 samples. To get a sense of how well the models are actually able to estimate this high dimensional psychometric test function, we include 4 random 2d slices through the 8 dimensional space after training on 500 random samples (Fig \ref{fig:2d_eval} \textbf{D}).

\subsection{Results on human behavioral data} 

To emphasize the generality and utility of our model, we evaluate performance on five participants from two real-world 6-dimensional datasets. All data are from visual psychophysical tasks. One dataset is a participant in a two-alternative forced choice (2AFC) task with 3000 trials from \cite{letham2022look}, and the remaining are four participants performing a 4 alternative forced choice (4AFC) from \cite{wuerger2020dataset} whose trial counts range from $\sim$200-500 depending on the subject\footnote{The former dataset is available at \url{https://github.com/facebookresearch/bernoulli_lse/tree/main/data}, and the latter at \url{https://www.repository.cam.ac.uk/handle/1810/304228}.}. In the 2AFC task, the participant is presented with an animated circular Gabor patch, one half of which has been scrambled to resemble white noise.  The scrambled side is selected at random. The stimulus varied along eight dimensions, six of which (contrast, background luminance, temporal and spatial frequency, size, and eccentricity) have data published. In the 4AFC task, a Gabor stimulus was presented in one of four quadrants of a screen, and participants were asked to select which quadrant contained the stimulus. This stimulus varied with size, orientation, frequency, and color. We used contrast as the intensity dimension for both tasks. For additional subject information and example stimuli, see supplementary materials.
We run 15-fold cross-validation, train our model on 80$\%$ of the data and test on $20\%$, and report cross-validated log-likelihood. We see across all five subjects that some variant of the semi-parametric model consistently has superior cross-validated log-likelihood on held-out trials, regardless of whether we use the MVN or exact semi-parametric variant. We also note that no one link function consistently performs best---this inconsistency across choice of link is one reason the specific parametric form of choice is still an area of active research. Nonetheless, we highlight that the semi-parametric model and MVN approximation are consistently strong performers.

\subsection{Active learning} 

Lastly, we evaluate the semi-parametric model's estimation of the 8-d test function using a variety of common Bayesian active learning schemes, including the BALD acquisition function based on mutual information \cite{houlsby2011bayesian,gardner2015bayesian,owen2021adaptive}, the BALV active learning scheme based on posterior variance \cite{settles2009active}, and GlobalMI, a global lookahead acquisition function based on threshold estimation \cite{letham2022look}. 
The computation of the GlobalMI aquisition function requires an MVN posterior on the latent function $\mathbf{z}$ and a probit link with a floor of 0, which means it can be applied directly with GP-RBF and GP-Linear, and to our MVN approximate model. To use GlobalMI with the full semi-parametric model, we switch the link to a probit with floor of 0 (from Gumbel with a floor set to chance), and apply the MVN approximation we derived above to the variational posteriors $q_c(\mathbf{f_c})$ and $q_k(\mathbf{f_k})$. In this setting we can still use the full semi-parametric model for evaluation. 
In addition to these baselines, we include the ThresholdBALV acquisition function as described earlier for use with the semi-parametric model.

\begin{figure*}
\centering
\includegraphics[width=1\textwidth]{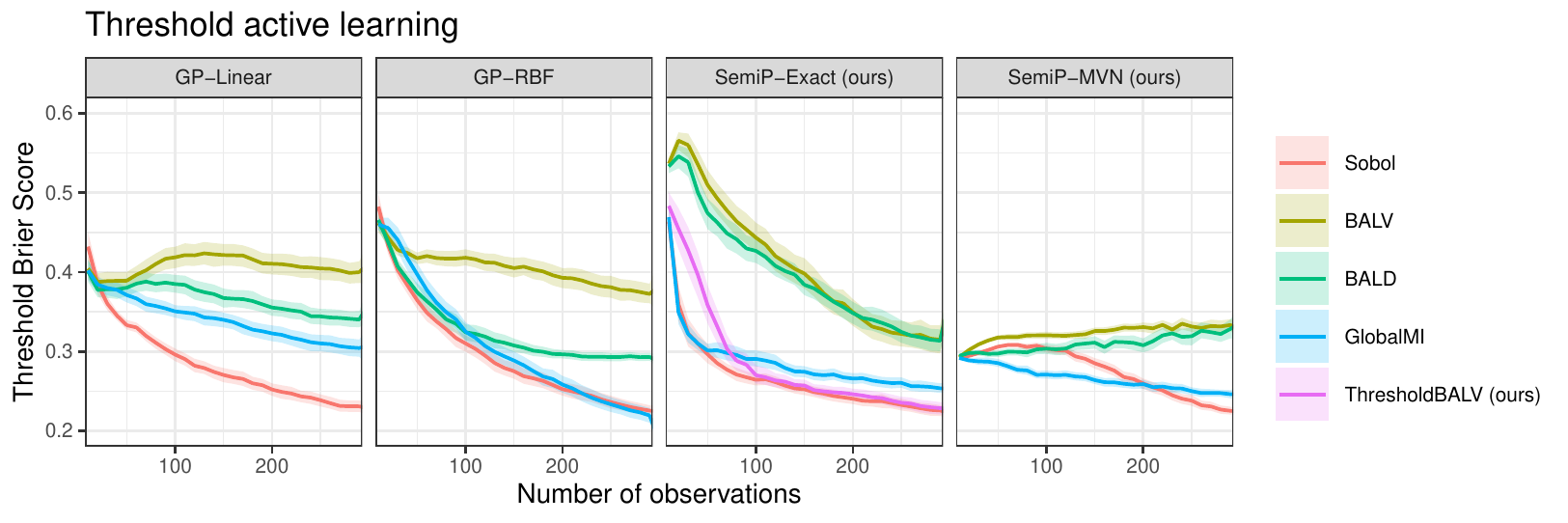}
\caption{Performance metrics for threshold estimation under a variety of common active learning paradigms. 
}
\label{fig:active_l}
\end{figure*}
Fig.~\ref{fig:active_l} shows performance of all models under the considered active learning schemes. Here, we are not using the standard Brier score as we did previously, which was over the full model posterior. Instead are using a Brier score on the sublevel-set (threshold) posterior, i.e.\ the Brier score on estimating the probability of $p$ above and below a threshold of 0.75. We choose this performance metric as the ThresholdBALV and GlobalMI acquition functions are based specifically on threshold estimation. Performance of these acquisition methods using other metrics are shown in the supplement. We see that the BALD and BALV baseline acquisition functions perform comparatively poorly, and this is true irrespective of the model. 
However, we see using threshold-based schemes (GlobalMI and ThresholdBALV) that the semi-parametric models perform well, especially during early acquisition. At the end of acquisition, GlobalMI acquisition in conjunction with the GP-RBF model performs marginally better than all other acquisition-based models. It is additionally important to note that quasi-random Sobol sampling for this 8-d function performs remarkably well against these active learning schemes, often as good or better than all acquisition functions tested. The unusual effectiveness of quasi-random sampling in this setting has been previously reported \cite{letham2022look} and we see it with our models as well as with the GP-RBF and GP-linear baselines. Exploring performance of these acquisition functions compared to quasi-random sampling for high dimensional psychophysics is an interesting avenue for future work.  Here, we simply wish to emphasize that the semi-parametric model and its MVN approximation are compatible and competitive under a variety of existing active learning schemes, and our proposed ThresholdBALV acquisition shows strong performance early on in sampling for an 8-d test function. For further evaluation of active learning for the semi-parametric models, see the supplement.  

\section{Conclusion}

We have demonstrated that a semi-parametric model for psychometric field estimation based on a parameterized sigmoid function can be adapted to high-dimensional psychophysical contexts using GPs as a non-parametric constraint on the sigmoid parameters. This semi-parametric approach not only offers parameters with scientific interpretation in the context of discrimination behavior, but offers accuracy improvements for estimating high dimensional tuning curves compared to competing methods. It does the latter both by providing a better prior, and by enabling a new active learning objective based on the semi-parametric functional form. We further introduce a moment-matching approximation to our model that can be used as a psychophysics-specific GP prior, or to produce approximate MVN posteriors compatible with analytic acquisition functions. We evaluate our contributions relative to baselines on both synthetic and real data and show a number of performance gains, especially for smaller sample sizes that are important in real human-in-the-loop experiments. 

\subsubsection{Limitations}

First, the evaluations in this paper either use synthetic test functions or human visual psychophysical data on a per-subject basis, so we cannot say at this stage how the model will perform with data from other sensory modalities or for cross-participant prediction. Second, we focus our evaluation on the Brier score in expectation over the posterior. While taking other metrics in expectation over the posterior shows similar behavior (as we demonstrate in the appendix), focusing on the posterior mean only (as done in prior work) changes the story somewhat. In particular, if we only consider the posterior mean of the RBF model, its key deficiency of having an overly flexible hypothesis space is mitigated and its performance looks stronger. Third, while we demonstrate that our work is compatible with active learning methods (even ones that require an MVN posterior), we do not offer an exhaustive evaluation of active learning methods, and the benefits of our contributions for active learning appear to be focused on small sample sizes. In line with this, we focus our evaluation and narrative around performance with relatively small data (consistent with the goal of sample-efficient psychophysics) but it is likely that with much larger datasets, the GP-RBF model's universality will let it match or outperform models with more restricted hypothesis spaces such as ours. 

\subsubsection{Ethics Statement}
Our work is primarily concerned with understanding low-level human perception, and as such carries relatively low risk of societal and ethical harm. Some risks include the misuse or de-anonymization of data, and overly broad or incorrect conclusions made based on data that is too limited, collected in a biased way, or based on misunderstanding or misusing the model. With respect to data misuse, we use only de-identified data that has been previously published, where informed consent was obtained, and is of low sensitivity (it is behavioral responses to simple visual stimuli). With respect to overly broad conclusions, we keep our claims narrowly focused on the quality of the model, and do not provide new interpretations or conclusions related to the datasets we use for evaluation. Furthermore, we think the specificity of our model for the psychophysics problem domain makes it less likely to be applied (and misused) in other settings than more generic models. On a more positive note, increasing sample efficiency for psychophysics studies may improve the experience of human research participants, who can sometimes be required to participate in dozens of hours of data collection when traditional grid or staircase methods are used.

\subsubsection{Computational Load}
With respect to computational load and environmental impact, the benchmarks were all carried out over the course of a few days ($\approx$70-100 hours) on a single EC2 \texttt{c6i.metal} node, and the cross-validation folds were performed over a smaller node over a similar period of time. These hours are largely taken up by replications across seeds (for benchmarks) and folds (for cross-validation)---for practical usage, the models we use take seconds to estimate on a typical laptop, which makes them accessible for use by most practitioners and researchers.


\section*{Appendix}
\appendix
\setcounter{equation}{0}
\setcounter{figure}{0}
\setcounter{table}{0}
\setcounter{page}{1}
\makeatletter
\renewcommand{\theequation}{A\arabic{equation}}
\renewcommand{\thefigure}{A\arabic{figure}}

\subsection{Information about 2d and 8d test functions}

For the 2d test function, the functional form across context and intensity are given by:
\begin{align*}
\theta_h(x_s) &= (-\frac12 \cos(0.6\pi x_s )+.55) * \left(-\frac12\sin( 0.3\pi x_s)+  \frac32\right), \\
f(x_i, x_s) &=\frac{10(x_i - \theta_h(x_s))}{2+\theta_h(x_s)}, 
\end{align*}

\noindent where $x_s$ and $x_i$ are the single stimulus and intensity dimensions. The output of $f$ is mapped through a probit to yield a probability value $p$. 

The 8d test function has a similar form, and is given by
\begin{align*}
\theta_h(x_{s^{(2)}}, \dots, x_{s^{(7)}}) =& \Big(-\frac{x_{s^{(2)}}}{2}  \big(\cos(0.6\pi x_{s^{(3)}} x_{s^{(4)}} + x_{s^{(5)}})+\frac{1}{2}\big) + x_{s^{(6)}} \Big)\cdot \\ &\Big(-x_{s^{(7)}}\sin(0.3\pi x_{s^{(3)}} x_{s^{(4)}} + x_{s^{(5)}})+( 2-x_{s^{(7)}})\Big) - 1,\\ 
f(x_i, x_{s^{(1)}}, \dots, x_{s^{(7)}}) =&\frac{x_i - \theta_h(x_{s^{(2)}}, \dots, x_{s^{(7)}})}{x_{s^{(1)}}\big(2+\theta_h(x_{s^{(2)}}, \dots , x_{s^{(7)}})\big)} ,
\end{align*}, 

where $x_{s^{(i)}}$ denotes the $i$th stimulus dimension and $f$ is put through a probit to yield probability $p$.

 \begin{figure*}[h!]
\centering
\includegraphics[width = .9\textwidth]{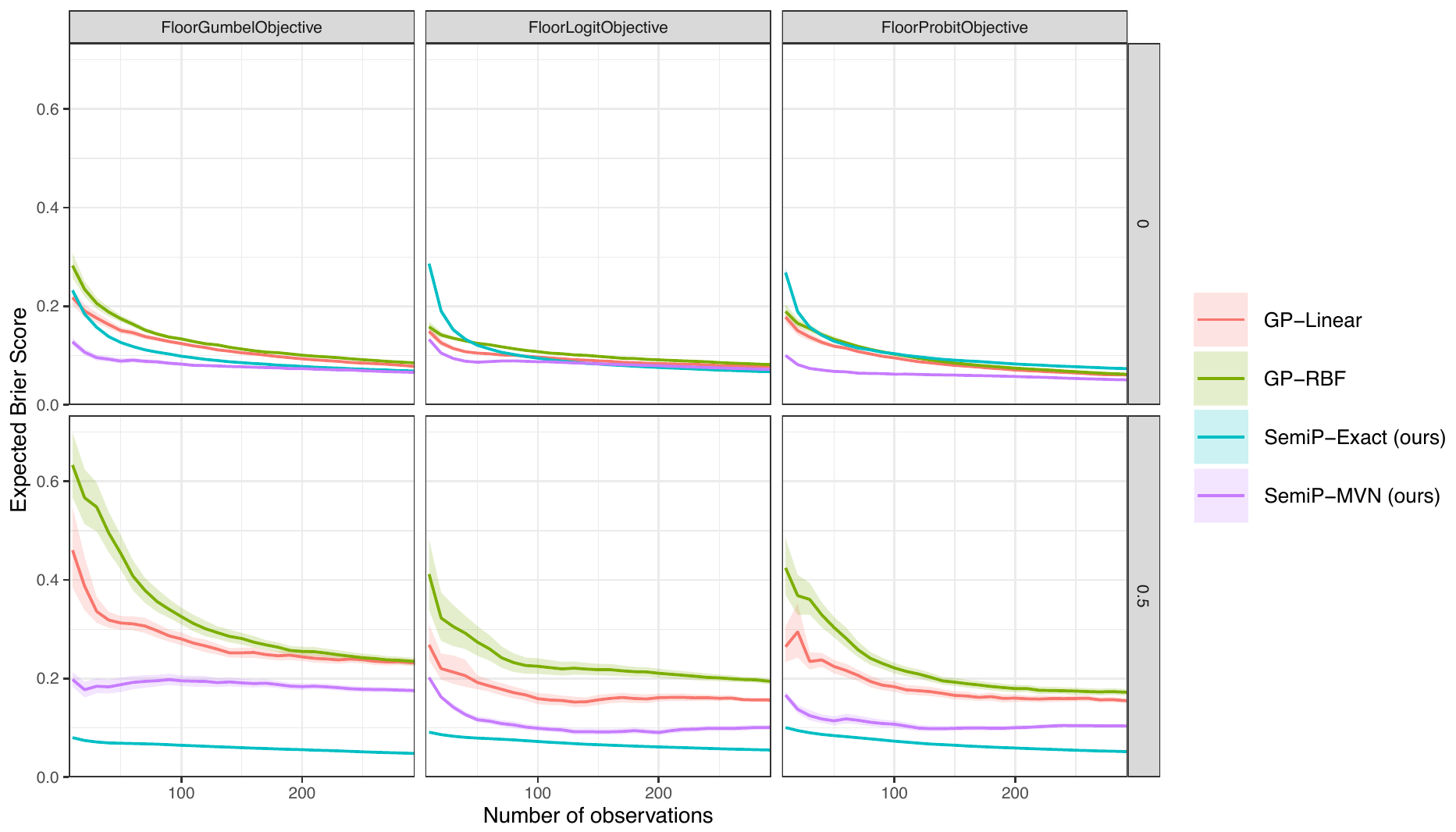}
\caption{Performance of all models on an 8d test function using all three links each with a floor of 0 and a floor of 0.5. }
\label{fig:link-floor}
\end{figure*}

\subsection{Analysis of the link function}

Figure~\ref{fig:link-floor} shows performance of all models on our 8d test function using the different link functions discussed in the manuscript (Gumbel, Logistic and Probit). We include versions of these links with a floor set to 0 and a floor set to the true floor of the test, 0.5. For the MVN models (GP-Linear, GP-RBF, and SemiP-MVN), adding the floor is harmful rather than useful -- this may be because the restricted output range created by the floor makes these models oversmooth the data. On the other hand, given a floor of zero, the choice of link for the MVN models is not important, with Probit performing slightly better than the others. In contrast, the exact semi-parametric model does benefit substantially from setting a nonzero floor based on domain knowledge of the problem, and the choice of link is again not particularly important, with Gumbel performing slightly better than the others. This result supports the claim that the full semi-parametric model is needed to take advantage of assumptions based on domain knowledge in this problem. 

\subsection{Alternate scoring metrics}

\begin{figure*}[htb]
\centering
\includegraphics[width = 1\textwidth]{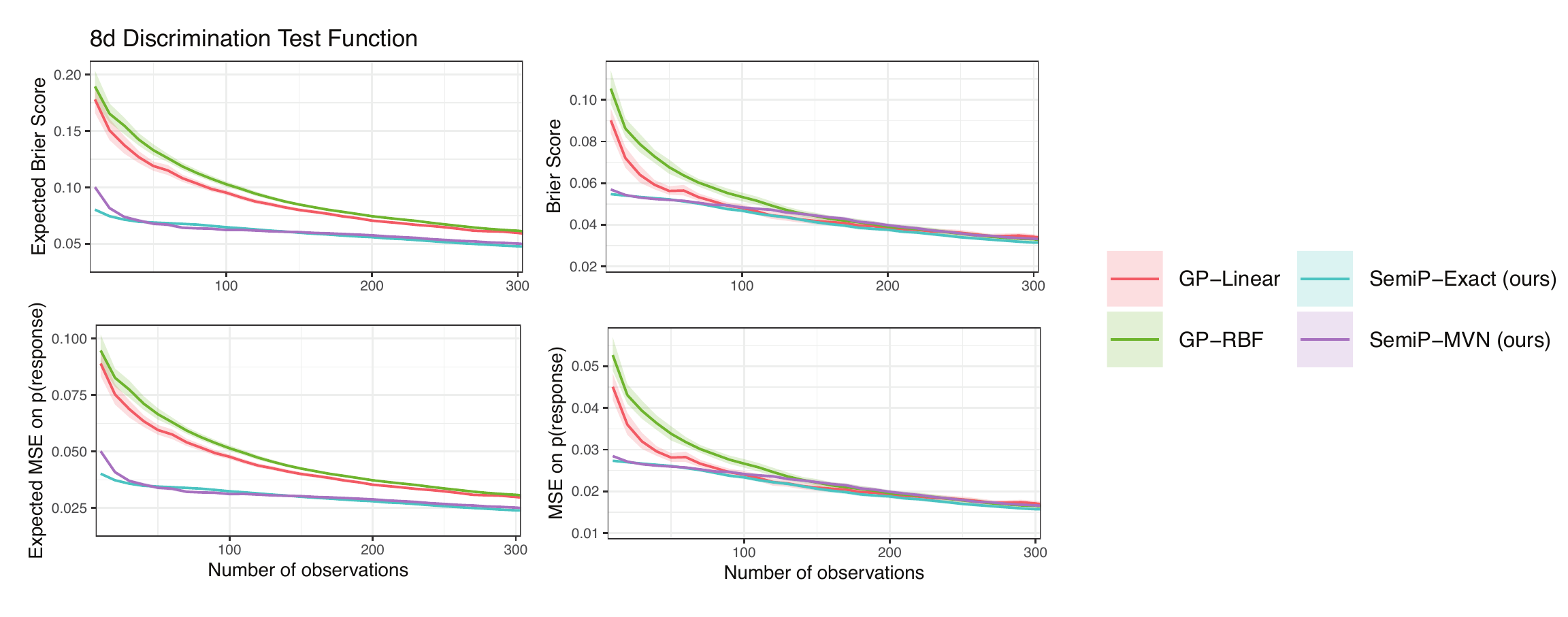}
\caption{Performance of the semi-parametric model on our 8d test function using the Brier score in expectation as reported in the main paper (top left), the Brier score with respect to the posterior mean (top right), mean squared error on the response probability, in expectation over the posterior (bottom left), and mean squared error on the response probability with respect to the posterior mean (bottom right). }
\label{fig:8d_measures}
\end{figure*}

 
 
 
As discussed in the main text, we chose to use the Brier score because it is a proper scoring rule (i.e.\ it measures how well-calibrated the posterior is) and we chose to use it in expectation over the GP posterior because it best reflects the quality of the whole posterior rather than just the posterior mean. Here we display alternate scoring metrics. We repeat performance measure of expected Brier score in the top left of Figure \ref{fig:8d_measures}. For consistency with prior work (e.g.\ of \citep{letham2022look}), the the top right panel of Figure~\ref{fig:8d_measures} shows the Brier score at the posterior mean, where the models are much closer to each other in performance, consistent with our claim that much of the benefit of the semi-parametric approach is reduced posterior uncertainty. However, we think a key benefit of GP modeling for both active learning and scientific utility is uncertainty quantification, and that therefore taking the metric in expectation is the better approach. When considering entirely different metrics, mean squared error (MSE) on the response probability looks substantially the same as the Brier score, with a larger benefit for the semi-parametric model when posterior uncertainty is taken into account (i.e.\ the bottom left of the figure looks like the top left, and the bottom right looks like the top right).

\subsection{Analysis of the hyperparameter $m$}

\begin{figure*}[htb]
\centering
\includegraphics[width = 1\textwidth]{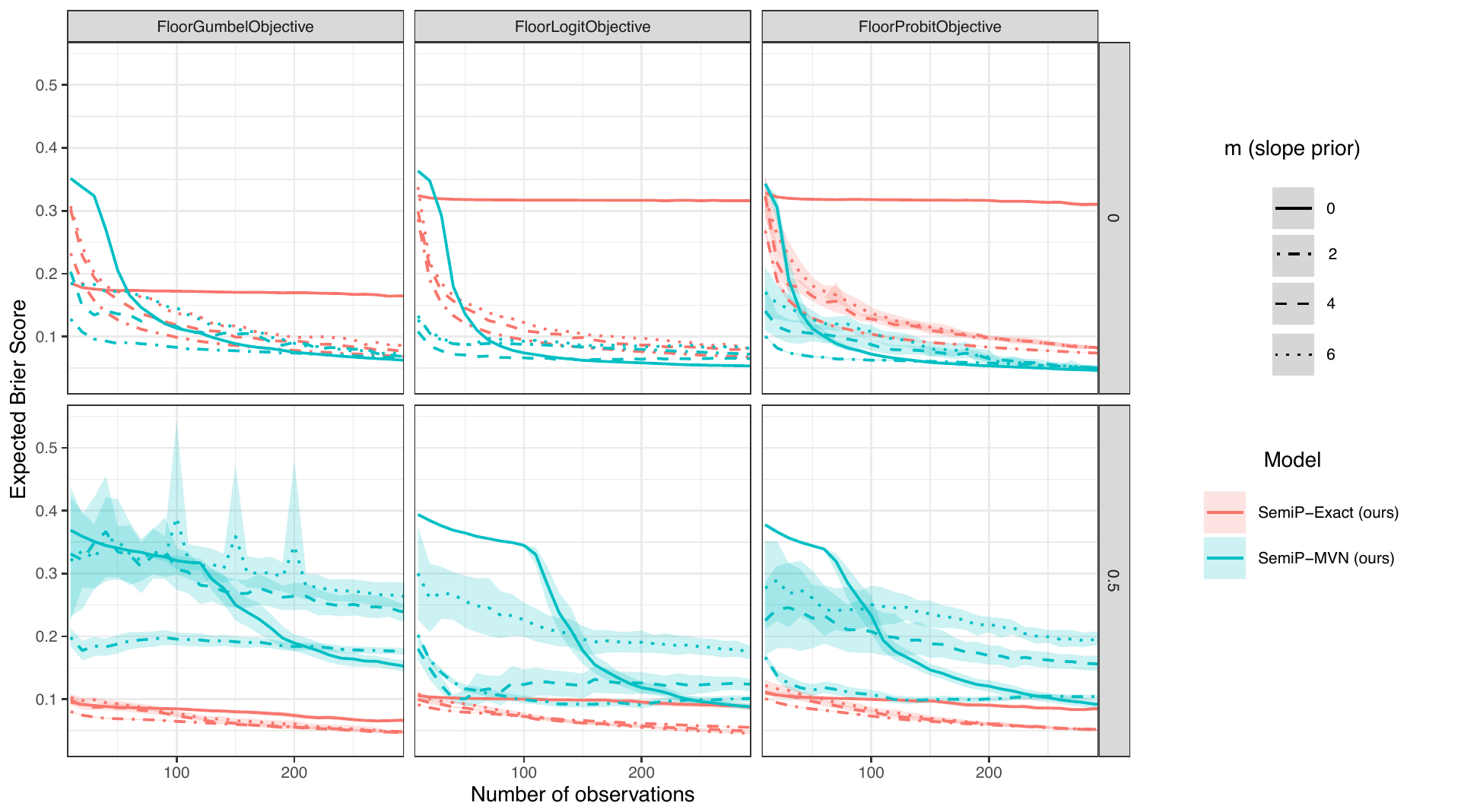}
\caption{Performance of the semi-parametric model on our 8d test function as we vary the GP slope mean, $m$}
\label{fig:m-plot}
\end{figure*}
 
In the main text, we set the mean of the semi-parametric mean slope GP to a constant value of 2, set based on manual exploration early in model development. Here we validate this choice by showing the performance of the exact semi-parametric model and MVN approximate semi-parametric model as we vary m values (0,2,4,6). As Figure~\ref{fig:m-plot} shows, generally the positive monotonicity prior created by a positive slope is highly beneficial for early performance of all models, but especially for the exact semi-parametric model. Outside of this, for the exact semi-parametric model the specific choice of this constant is generally unimportant. In the MVN approximate model the choice of our slope GP mean has a greater affect on performance, especially when the floor is set to 0.5. However, for our purposes we set $m=2$ early on, and continued with that value throughout the manuscript, including on real data. No re-tuning of this value was done to optimize semi-parametric performance. 

\begin{figure*}[htb]
\centering
\includegraphics[width = 1\textwidth]{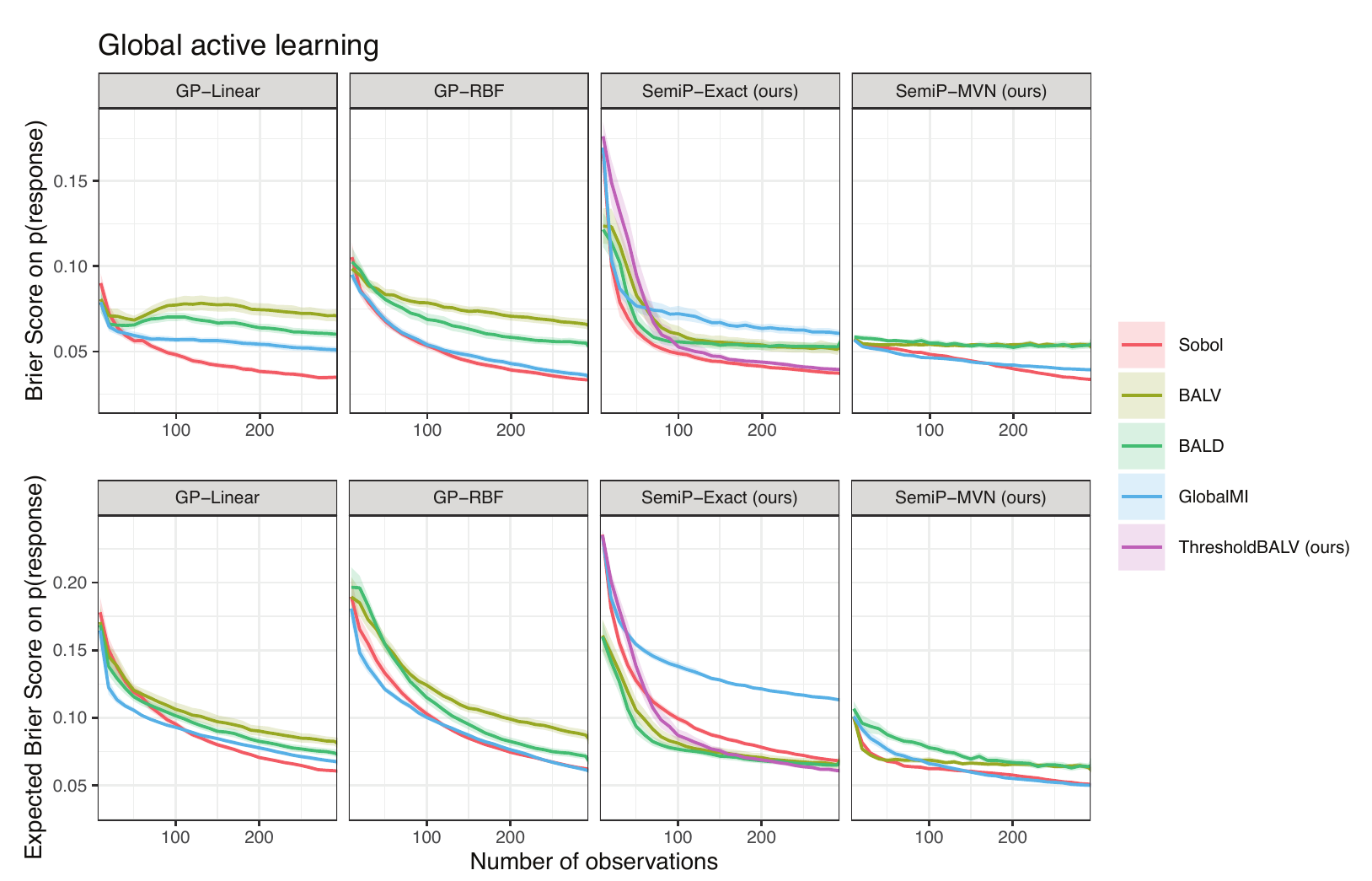}
\caption{Performance of active learning methods using different metrics}
\label{fig:al_supp}
\end{figure*}

\subsection{Derivation of the factorized ELBO}

The standard evidence lower bound can be written as \cite{Hensman2015b}:
$$
\mathcal{L}:=\mathbb{E}_{q(\mathbf{f})}[\log p(\mathbf{y} \mid \mathbf{f})]-\mathrm{KL}[q(\mathbf{f}) \| p(\mathbf{f})]
$$
Where here, $q(\mathbf{f}) = q(\mathbf{f_k})q(\mathbf{f_c})$ and $p(\mathbf{y}|\mathbf{f}) = p(\mathbf{y}|\mathbf{f_k}, \mathbf{f_c}$) and $p(\mathbf{f}) = p(\mathbf{f_k})p(\mathbf{f_c})$. For notational simplicity, let, $\mathbf{f_k} = \textbf{x}$, $\mathbf{f_c} = \textbf{y}$ and  So the second term in the ELBO can be written 
\begin{align*}
\mathrm{KL}[q(\mathbf{x})q(\mathbf{y}) \| p(\mathbf{x})p(\mathbf{y})) = &\iint p(\mathbf{x})p(\mathbf{y}) \log{\frac{p(\mathbf{x})p(\mathbf{y})}{q(\mathbf{x})q(\mathbf{y})}} d\mathbf{x}d\mathbf{y}\\
= \iint p(\mathbf{x})p(\mathbf{y})\big(\log(p(\mathbf{x})) &+ \log(p(\mathbf{y})) \\ - \log(q(\mathbf{x})) -& \log(q(\mathbf{y}))\big)d\mathbf{x}d\mathbf{y}\\
= \iint  p(\mathbf{x})p(\mathbf{y})\Big(\log\frac{p(\mathbf{x})}{q(\mathbf{x})} &+ \log\frac{p(\mathbf{y})}{q(\mathbf{y})}\Big)d\mathbf{x}d\mathbf{y}\\ 
=\int p(\mathbf{x})\log\frac{p(\mathbf{x})}{q(\mathbf{x})}d\mathbf{x} &+ \int p(\mathbf{y})\log\frac{p(\mathbf{y})}{q(\mathbf{y})}d\mathbf{y} \\ 
= \mathrm{KL}[q(\mathbf{x}) \| p(\mathbf{x})] &+ \mathrm{KL}[q(\mathbf{y}) \| p(\mathbf{y})]
\end{align*}
We use the above form in our evidence lower bound expression in the main text in equation 4.

\subsection{Additional hyperparameter details}
In addition to the hyperparameters explored above, we report additional reproducibility details: 
\begin{itemize}
    \item We used 100 inducing points taken from a Sobol sequence at which we approxiamted the variational posterior. 
    \item For both slope and intercept GPs, we used a Radial Basis Function (also known as Squared Exponential) kernel, with independent $Gamma(3, 6)$ priors on lengthscales (a form of Automatic Relevance Determination) and $Gamma(1.5, 1.0)$ prior on the variance. 
\end{itemize}

\subsection{Further exploration of active learning}

Here, we evaluate our active learning contributions by other metrics. Even though ThresholdBALV and GlobalMI are intended for threshold estimation, we can still evaluate them using global metrics such as the expected Brier score, and we do in Figure~\ref{fig:al_supp}. Consistent with results elsewhere in the text, our models perform well (and SemiP-MVN performs best). 

\subsection{Additional dataset detail and example stimuli}
\begin{figure*}[htbp]
\centering
\includegraphics[width=0.5\textwidth]{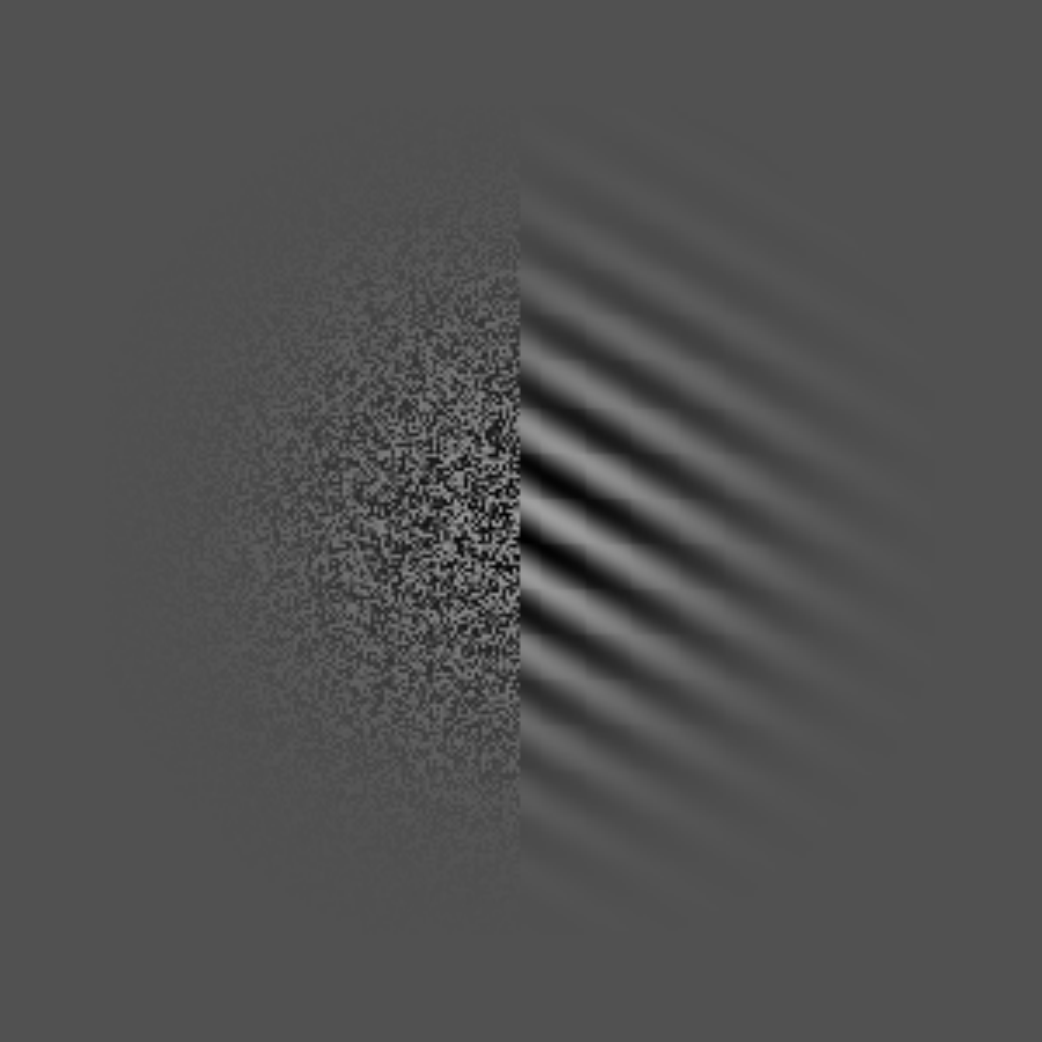}
\caption{Example stimulus from Letham et al.\ 2022 Fig. 4, licensed by MIT license, retrieved from \texttt{https://github.com/facebookresearch/bernoulli\_lse/blob/main/figures/pdfs/p100.png}}
\label{fig:letham_stim}
\end{figure*}

2AFC data studying the contrast sensitivity function (CSF) was taken from \cite{letham2022look} and is available at \texttt{https://github.com/facebookresearch/bernoulli\_lse/}. Stimuli were animated sinusoidal gratings (aka Gabor patches), with varying luminance, spatial and temporal frequencies, and contrast. Stimulus configurations were drawn from a quasi-random sequence over the input domain and a participant judged which side of the stimulus had been scrambled. An example stimulus is shown in Fig.~\ref{fig:letham_stim}. 

\begin{figure*}[htbp]
\centering
\includegraphics[width=\textwidth]{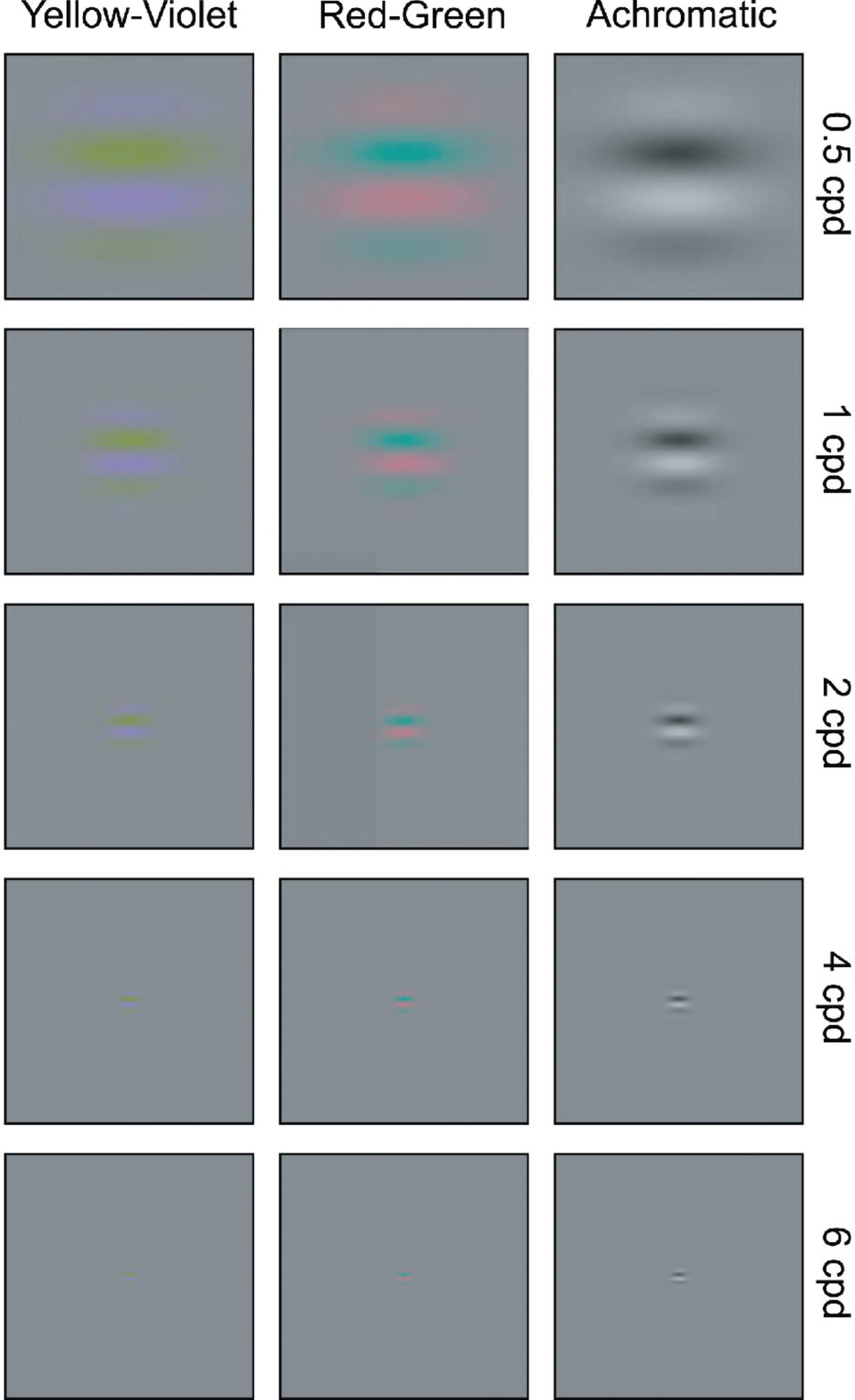}
\caption{Example stimuli from Wueger et al. 2020, licensed CC-BY (license available at \texttt{https://creativecommons.org/licenses/by/4.0/}. Image retrieved from \texttt{https://jov.arvojournals.org/article.aspx?articleid=2765519}, Fig.\ 4}
\label{fig:wuerger_stims}
\end{figure*}

The 4AFC dataset is from \cite{wuerger2020spatio}, also studying CSFs, and was requested from the authors. Stimuli consisted of Gabor patches of different spatial frequencies, angular sizes, background luminance, and contrast. Participants judged which quadrant of the screen the stimulus appeared in. In both cases contrast was the intensity dimension and the remaining dimensions were context. We used data from four participants with a comparatively small number of trials. Specifically, we used subjects 10, 11, 13, and 19 with 508, 477, 201 and 318. Example stimuli are shown in Fig.~\ref{fig:wuerger_stims}. 

\clearpage
\bibliographystyle{unsrt} 
\bibliography{mike}

\end{document}